\documentclass[fleqn,usenatbib,useAMS]{mnras}

\usepackage{graphicx}	
\usepackage{amsmath}	
\usepackage{amssymb}	
\usepackage{multicol}        
\usepackage{bm}		
\usepackage{pdflscape}	
\usepackage{xcolor,ulem}
\usepackage{adjustbox}

\newcommand{\be}{\begin{equation}}
\newcommand{\ee}{\end{equation}}
\newcommand{\ba}{\begin{eqnarray}}
\newcommand{\ea}{\end{eqnarray}}

\newcommand\restr[2]{{
  \left.\kern-\nulldelimiterspace 
  #1 
  \vphantom{\big|} 
  \right|_{#2} 
  }}
  
\definecolor{blazeorange}{rgb}{1.0, 0.4, 0.0}
\definecolor{seagreen}{rgb}{0.18, 0.55, 0.34}
\definecolor{rufous}{rgb}{0.66, 0.11, 0.03}
\definecolor{royalfuchsia}{rgb}{0.79, 0.17, 0.57}
\definecolor{scarlet}{rgb}{1.0, 0.13, 0.0}
\definecolor{royalpurple}{rgb}{0.47, 0.32, 0.66}
\definecolor{darkblue}{rgb}{0, 0, 0.66}

\usepackage[T1]{fontenc}
\usepackage{ae,aecompl}
\usepackage{newtxtext,newtxmath}

\usepackage{ulem}

\title[Sw J1644 as an off-axis jet]{Swift J1644+57 as an off-axis Jet}

\author[P. Beniamini, T. Piran and T. Matsumoto]{
Paz Beniamini$^{1,2}$\thanks{E-mail: pazb@openu.ac.il},
 Tsvi Piran$^{3}$ and Tatsuya Matsumoto$^4$
\\
$^{1}$Department of Natural Sciences, The Open University of Israel, P.O Box 808, Ra'anana 4353701, Israel\\
$^{2}$Astrophysics Research Center of the Open university (ARCO), The Open University of Israel, P.O Box 808, Ra'anana 4353701, Israel\\	
$^3$Racah Institute of Physics, the Hebrew University, 91904, Jerusalem, Israel\\
$^4$Department of Physics and Columbia Astrophysics Laboratory, Columbia University, Pupin Hall, New York, NY 10027, USA}

\pubyear{2023}
\begin{document}
\label{firstpage}
\pagerange{\pageref{firstpage}--\pageref{lastpage}}
\maketitle

\begin{abstract}
One of the intriguing puzzles concerning  Swift J1644+57, the first jetted tidal disruption event (TDE) discovered, is the constant increase in its jet energy, as implied by radio observations. During the first two hundred days the jet energy has increased by an order of magnitude. We suggest that the jet was viewed slightly off-axis. In this case, the apparent energy increase arises due to the slowing down of the jet and the corresponding broadening of its beaming cone. Using equipartition analysis, we infer an increasing jet energy as a larger region of the jet is observed. A simple off-axis model accounts nicely for the multi-wavelength radio observations, resolving this long-standing puzzle. The model allows us to self-consistently evolve the synchrotron signature from an off-axis jet as a function of time. It also allows us to estimate, for the first time, the beaming angle of the jet, $\theta_0 \approx 21^{\circ}$. Considering existing limits on the black hole mass, $\lesssim 10^{7}M_{\odot}$, this angle implies that the prompt phase beaming corrected luminosity of Swift J1644+57, $\sim 10^{47}$ergs/sec, was super Eddington. We also present a closure relation between the spectral and temporal flux for off-axis jets, which can be used to test whether a given radio transient is off-axis or not.
\end{abstract}



\section{Introduction}
\label{sec:Intro}
Swift J1644+57 (hereafter, Sw J1644) was the first identified jetted tidal disruption event (TDE). It was first detected in X-rays \citep{Burrows2011,Bloom2011,Levan+2011}. 
Its location in the center of the host galaxy suggested that it was a TDE. The observed declining rate of the X-ray luminosity is also consistent with the characteristic mass fallback rate in TDEs. The observed luminosity, which was orders of magnitude larger than Eddington, suggested that the emission must have been jetted. While Sw J1644 was never observed in the optical band, most likely due to severe dust extinction, it had a radio signal \citep{Zauderer2011} that was interpreted \citep{Berger2012,Metzger+2012} as the radio afterglow produced by the interaction of the jet with the surrounding matter. Surprisingly, equipartition analysis revealed that the energy within the radio-emitting region increased continuously on a time scale of 200 days \citep{Barniol2013A,Eftekhari2018,Cendes2021}. A similar result was obtained using afterglow modeling \citep{Berger2012,Zauderer2013}.

Several models have been suggested to explain the puzzling energy increase. \cite{Berger2012} proposed energy injection by the central engine. 
However, as we discuss below this is at odds with the X-ray light curve.
\cite{Mimica2015} and \cite{Generozov2017} suggested  that the underlying jet had an angular structure, whereby the core fast component (with $\Gamma\approx 10$) is surrounded by a slower ($\Gamma\approx 2$) more energetic sheath that is slower to decelerate and therefore adds to the observed emission and inferred energy only at late times. While the origin of the injected energy is different, modeling of the afterglow, in this case, is rather similar to the models used by \cite{Berger2012}.
A very different possibility explored by 
\cite{Kumar2013} is that inverse Compton (IC) losses suffered by the synchrotron radiating electrons in the forward shock  (due to IC scattering of the  X-rays), could provide a flattening of the radio and IR lightcurve without a need to inject energy to the blast-wave at late time. In this model, all the energy is injected at once but due to significant IC cooling by the X-ray photons only a small fraction of the electrons' energy is emitted via synchrotron at radio wavelengths. As the X-rays flux decreases, this effect diminishes and a larger fraction of the energy is emitted in the radio. 

Recently \cite{Matsumoto2023} developed a formalism for the equipartition analysis of relativistic off-axis `top-hat' jets (i.e. jets with a roughly uniform energy per solid angle within their core and negligible energy outside of it). This formalism introduces an additional free parameter to the relativistic equipartition analysis of \cite{BarniolDuran2013}, the viewing angle of an observer from the radio-emitting region $\theta_{\rm eq}$. A generic feature of a relativistic off-axis solution is that as the source slows down and its Lorentz factor decreases an off-axis observer sees a larger fraction of the source. This results in an increase of the apparent\footnote{The equipartition analysis estimates the energy within the observed region.}  energy inferred from the observations. A characteristic feature of emission from a `top-hat' like jet is a rapid increase in the flux at a given frequency, as seen for example in AT 2018hyz, whose radio light curve reveals a $\propto t^5$ increase of the peak flux \citep{Cendes+2022b}. We emphasize that if the jet has significant kinetic energy at latitudes beyond the core, the observed off-axis signal can be very different \citep{Rossi02,KG2003,Eichler&Granot2006}. In particular, the rise of the signal during the phase described above can become more shallow (and one can relate between the angular energy distribution and the rate at which the flux is evolving, see \citealt{GG2018,BGG2020,Takahashi2021}). Indeed this is the commonly accepted picture in the case of GRB 170817A, the first GRB associated with a GW detected BNS-merger \citep[e.g.,][]{Lazzati2018,Margutti2018,Ghirlanda2019,Hotokezaka+19,Troja2019,Gill+19}. Furthermore, depending on the viewing angle of the observer, off-axis signals from such structured jets can manifest in either single or double peaked light-curves (see \citealt{BGG2020,BGG2022} for details).

Motivated by the off-axis equipartition analysis we explore here the possibility that Sw J1644 was viewed off-axis and the apparent increase of the energy implied by the radio observations arose when it was slowing down. The peak at around 200 d was reached when we observed the whole jet. Applying Occam's razor, we seek the simplest solution of this type and ask whether it might provide a good fit to observations. As such we consider in our analysis a toy model consisting of a simple `top-hat' jet and do not add additional free parameters describing the exact profile of the kinetic energy and Lorentz factor of the jet as a function of latitude. The structure of the paper is as follows. We begin in \S \ref{sec:equi} with a description of the results of the off-axis equipartition analysis of the radio observations of Sw J1644. In \S \ref{sec:FwdModel} we describe a blast wave propagation model for the multi-wavelength flux of an off-axis jet. Using this formalism we analyze in \S \ref{sec:SwJ1644model}  the radio and X-ray observations of Sw J1644 and using MCMC we obtain a best-fit set of parameters. We summarize our findings and conclude in \S \ref{sec:conc}.

\section{Off-axis equipartition analysis } 
\label{sec:equi}

\begin{figure}
\begin{center}
\includegraphics[width=75mm, angle=0]{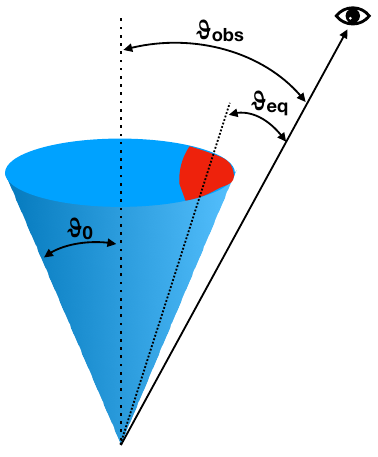}
\caption{The overall geometry. The jet opening angle is $\theta_0$.  The observer is at an angle $\theta_{\rm obs}$ to the center of the jet. The momentary observed region is marked in red and the observer is at an angle $\theta_{\rm eq}$ from the center of this region. The size of the red region increases with time as the jet slows down and with this $\theta_{\rm eq}$ increases until it reaches $\theta_{\rm obs}$ when the majority of the jet is observed.}
\label{fig:geometry}
\end{center}
\end{figure}

\begin{figure}
\begin{center}
\includegraphics[width=85mm, angle=0]{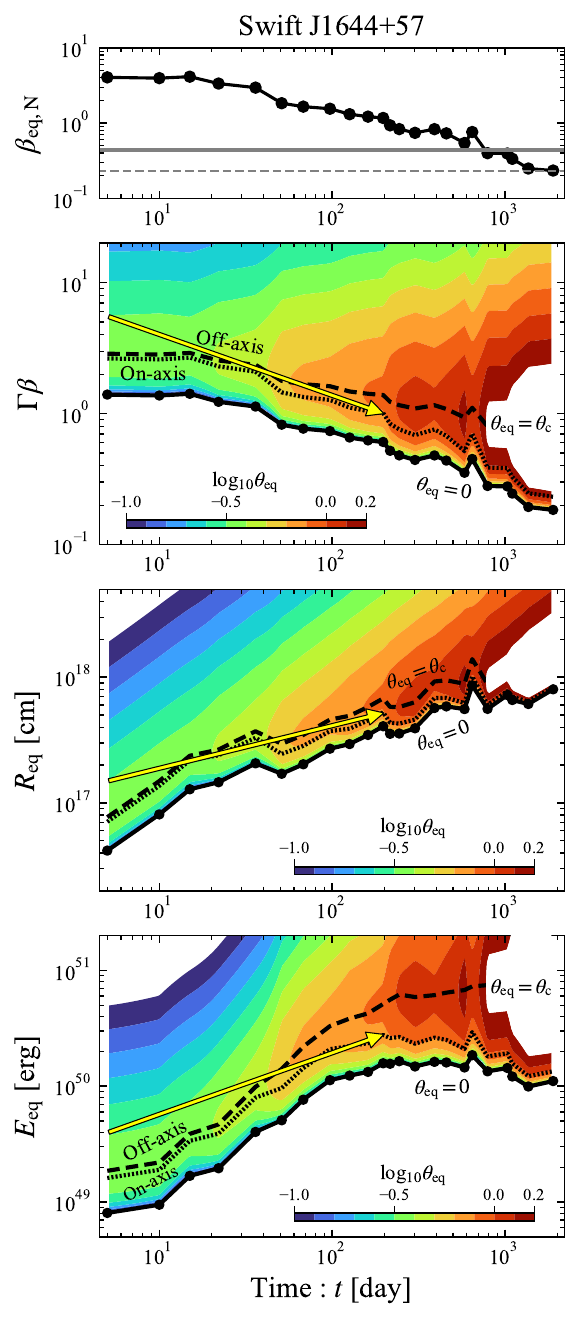}
\caption{The time evolution of the parameter $\beta_{\rm eq,N}$, the emitting region four-velocity $\Gamma\beta$, and the equipartition radius $R_{\rm eq}$ and energy $E_{\rm eq}$ for the jetted TDE Swift J1644+57. In the top panel, the solid (dashed) gray line shows the critical value of $\beta_{\rm eq,N}=0.44$ ($0.23$) above which there is no forbidden region in $\Gamma\beta$. In the second to fourth panels, the quantities take a minimum at $\theta_{\rm eq}=0$. The corresponding viewing angles are shown by color contours. The black dashed and dotted curves show the contours for $\theta_{\rm eq}=\theta_{\rm c}$ (maximal viewing angle - see text) and the boundary between 
on and off-axis ($\theta_{\rm eq}\simeq1/\Gamma$). Yellow arrows show a possible time evolution of the radio source, from off to on-axis viewed configurations. 
Observables are linearly interpolated to depict the color map. }
\label{fig:result_j1644}
\end{center}
\end{figure}

We analyze the radio observations of Sw J1644 using the generalized equipartition formalism developed in \cite{Matsumoto2023} that allows for the possibility of an off-axis observer. This adds an extra degree of freedom, the {\it momentary} viewing angle, $\theta_{\rm eq}$, which is defined as the angle between the observer line of sight and the center of the {\it observed} region (see Fig. \ref{fig:geometry}). Because of the nature of the equipartition analysis that deals only with the observed region, we have $\theta_{\rm eq} \le \theta_{\rm obs}$. Namely, $\theta_{\rm eq}$ is smaller or equal to the angle between the center of the jet and the direction towards the observer, and it varies with time.  The observables used in the analysis are: $\nu_{\rm p}$, $F_{\rm p}$, $z$, and $t$, the peak frequency and the flux density at the peak, the redshift of the source (and the corresponding luminosity distance, $d_{\rm L}(z)$), and the observation time.  

A critical parameter in this analysis is the apparent velocity:  
\begin{align} \beta_{\rm eq,N}&\equiv\frac{(1+z)R_{\rm eq,N}}{ct}    \label{eq:beta}\\
&\simeq0.73\,\left[\frac{F_{\rm p,mJy}^{\frac{8}{17}}d_{\rm L,28}^{\frac{16}{17}}\eta^{\frac{35}{51}}}{\nu_{\rm p,10}(1+z)^{\frac{8}{17}}}\left(\frac{t}{100\,\rm day}\right)^{-1}\right]f_{\rm A}^{-{7}/{17}}f_{\rm V}^{-{1}/{17}} \ ,
	\nonumber
\end{align}
where  $c$ is the speed of light and $R_{\rm eq,N}$ is the Newtonian equipartition radius  \citep[see][]{Matsumoto2023}.
Following \cite{BarniolDuran2013} we define three dimensionless quantities:  $\eta =1$ if $\nu_{\rm a} > \nu_{\rm m} $ and $\eta =\nu_{\rm m}/\nu_{\rm a}$ otherwise ($\nu_{\rm a}$ and $\nu_{\rm m}$ are the observationally derived synchrotron self-absorption and characteristic frequencies),  $f_{\rm A}\equiv A/\left(\pi R^2/\Gamma^2\right)$ and $f_{\rm V}=V/\left(\pi R^3/\Gamma^4\right)$
are the area and volume filling factors respectively with $R$ and $\Gamma$ the distance from the origin and the Lorentz factor respectively. These filling factors are of order 
unity\footnote{Here and elsewhere in this paper we consider a one zone model. Variations within the source, such as patchiness, break this assumption and may modify the results. } 
when the observed region's size is $\sim1/\Gamma$. Finally, we use the convention $Q_x = Q/10^x$ (cgs) except for the flux density $F_{\rm p,mJy}=F_{\rm p}/{\rm mJy}$.

The parameter $\beta_{\rm eq,N}$ describes an apparent velocity of the emitting source. Importantly, a critical criterion is whether $\beta_{\rm eq,N}\lessgtr0.23$.  Only if $\beta_{\rm eq,N}>0.23$  can the solution transit from the relativistic off-axis branch to the Newtonian on-axis branch. Such a transition is essential in any realistic afterglow model in which the jet slows down and eventually becomes Newtonian. Note that the critical value corresponds to the maximal viewing angle of $\theta_{\rm eq}=\pi$. However, realistically the jet symmetry restricts us to a narrower angle range $\theta_{\rm eq}<\pi/2$. In this case, we find that the critical value becomes larger $\beta_{\rm eq,N}\simeq0.44$. Below this value, the relativistic and Newtonian branches split into two disconnected parts. Therefore, the transition from off-axis to on-axis view should happen only when  $\beta_{\rm eq,N}>0.44$.
The top panel of Fig.~\ref{fig:result_j1644} depicts the evolution of $\beta_{\rm eq,N}$ (data is taken from \citealt{Eftekhari2018}). We set the {origin of} time as the time of the first detection. As $\beta_{\rm eq,N} \ge 0.44$ and it approaches the critical value $0.44$,  an off-axis solution is viable, as off-axis to on-axis transition is possible. 

For each observation epoch, we derive   \citep{Matsumoto2023}:
a possible range of four-velocities, equipartition radii, and energies as a function of the momentary viewing angle $\theta_{\rm eq}$ (shown in Fig.~\ref{fig:result_j1644}).  The minimal values for all three quantities are obtained, as expected, for an on-axis observer with $\theta_{\rm eq}=0$. These values for $\theta_{\rm eq}=0$ are consistent with those of \cite{Barniol2013A} and \cite{Eftekhari2018}.\footnote{In our estimate of $E_{\rm eq}$ we included only the energies of the non-thermal electrons and magnetic fields and we did not include additional energy such as the energy of the protons and possible deviation from the equipartition, which are considered by those authors. } 
Off-axis solutions with $1/\Gamma<\theta_{\rm eq}$ are allowed for different viewing angles. However, $\theta_{\rm eq}$ is capped from above by the critical angle $\theta_{\rm eq}<\theta_{\rm c}\simeq\beta_{{\rm eq,N}}^{-17/24}$. 
There is no solution that minimizes the energy for $\theta_{\rm eq}>\theta_{\rm c}$  \citep[see][for details]{Matsumoto2023}. We show the corresponding viewing angles by color contours. To depict a continuous color map, we linearly interpolate the observation data between successive observation epochs. The decreasing velocity parameter $\beta_{\rm eq,N}$ leads to an increase of the momentary maximal viewing angle and a decrease of the minimal four-velocity (obtained for $\theta_{\rm eq}=0$). As discussed above, once $\beta_{\rm eq,N}<0.44$, the relativistic and Newtonian branches become disconnected. 

There are two possible geometries: on and off-axis. Each one leads to a different temporal evolution of the source. If the jet is viewed on-axis, as assumed in previous works, energy has to be injected into the jet (see however \citealt{Kumar2013}).  Another possibility is that the jet is viewed initially off-axis.  
Both options are shown in Fig.~\ref{fig:result_j1644} in which the allowed solutions are divided into on-axis and off-axis regimes. In particular, the figure depicts (with a yellow arrow)  a (schematic) possible off-axis solution.\footnote{We stress that the evolution shown by the yellow arrow is just a simplified example. A realistic evolution would not track the straight line in the figure, and in particular will never intersect with the on and off-axis boundary multiple times.} At the beginning of the observation, as the jet is viewed slightly off-axis the emitting region is just the edge of the jet. As time increases, the jet decelerates and contributes to the observed emission. Therefore, both the momentary viewing angle and the inferred equipartition energy increase. The flattening of the energy at $\simeq200$ days suggests that the whole jet has entered our line of sight, by that stage.
Motivated by this result, we explore below a self-consistent model of the dynamics and radiation of a relativistic forward shock afterglow, as viewed by an off-axis observer. We will show that such a model can account for the observations of Sw J1644.

\section{The off-axis jet model}
\label{sec:FwdModel}
To avoid adding unnecessary free parameters, we consider a `top-hat' jet in which $dE_{\rm k}/d\Omega=E_{\rm k,iso}/4\pi$ is taken to be constant up to some opening angle $\theta_0$. The jet moves initially  relativistically with $\Gamma_0\gg \theta_0^{-1}$.
The jet propagates into an external medium with a mass density profile $\rho =A r^{-k}$ with $0\leq k \leq 2.5$ \footnote{ We note that the real environment of supermassive black-holes may be more complex than a simple PL profile (see e.g. \citealt{Quataert2004}). However, without a specific model for $k(r)$, this will introduce additional free parameters that cannot be well constrained by the available data of Sw J1644. We also point out that as will be shown below, in our best fit for Sw J1644, the radius of the blast-wave changes only by a factor of $\sim 10$ between the earliest and latest afterglow observations (see figure \ref{fig:radius}), so introducing a complex behavior of $k(r)$ within this range is unlikely to affect the fit by any significant amount relative to the `averaged’ PL behavior used in our analysis.}.
It is convenient to recast $A$ in terms of the number density at some fixed reference radius, $A=m_{\rm p} n(R_0) R_0^{k}$, where $m_{\rm p}$ is the proton mass. We use a reference radius of $R_0=10^{18}\mbox{ cm}$.
As the jet propagates it is decelerated by, and drives a forward shock through, this external medium. The deceleration radius is given by 
\begin{equation}
    R_{\rm dec}=\left[\frac{(3-k)E_{\rm k,iso}}{4\pi Ac^2\Gamma_{\rm 0}^2}\right]^\frac{1}{3-k} \ .
\end{equation}
For an on-axis observer, deceleration happens at  $t_{\rm dec}=(1+z)R_{\rm dec}/(2c\Gamma_0^2)$.
After this time, the jet slows down, while conserving the energy in the shocked region, i.e. $\Gamma=\Gamma_0(t_{\rm on}/t_{\rm dec})^{\frac{k-3}{8-2k}}$ (where $t_{\rm on}$ refers to an observation time for an observer viewing along the jet core, i.e., on-axis).

This post-deceleration evolution persists at least until the jet-break time, $t_{\rm jb}=t_{\rm dec}(\theta_0 \Gamma_0)^{\frac{8-2k}{3-k}}$, at which the jet satisfies the condition $\Gamma(t_{\rm jb})=\theta_0^{-1}$ (where for clarity we have assumed here $\theta_0\ll 1$). At this point, we can consider two limiting cases for the dynamics. The first possibility is fast lateral spreading, which is motivated by both theoretical \citep{Rhoads1999,SariPiran1999,KG2003,Wygoda2011} and observational works (e.g. from the post-peak temporal decline in the afterglow of GRB  170817A; \citealt{Margutti2018,Troja2020}) and which is generally considered to provide a good approximation of `top-hat' like jets with sufficiently narrow cores. Here, the proper velocity can be assumed to decrease exponentially in time in the jet engine frame \citep{Rhoads1999} after this point. Since, for a source moving relativistically towards the observer, the arrival time, $t_{\rm ar}$, is proportional to $t_{\rm em}/\Gamma(t_{\rm em})^2$ (where $t_{\rm em})$ is the emission time), a rapid decrease of $\Gamma$ in the jet engine frame corresponds to a much longer apparent duration in the observer frame, (lasting approximately $t_{\rm jb}\theta_0^{-2}$) during which the jet had little time to propagate forwards. As a result, the radius of the outflow can to a good approximation be assumed to stall at $t_{\rm jb}$, as the flow starts spreading sideways at an order unit fraction of the speed of light. 
Recalling that in the (on-axis) observer frame $t_{\rm on}\propto R/\Gamma^2$, this leads to $\Gamma\propto t_{\rm on}^{-1/2}$ and to the jet opening angle evolving as $\theta(t_{\rm on})\approx \Gamma(t_{\rm on})^{-1}$.
The second limiting case is that of no lateral spreading. This too is motivated by both theoretical and observational considerations \citep{Granot2001,vanEerten2011,Granot&Piran2012,Berger2012} and has been argued to be more appropriate for the case of TDE jets \citep{Matsumoto&Piran2021b,Matsumoto&Metzger2023}.
It will, in general be expected to provide a better explanation for wider jets and / or jets with significant lateral structures \citep{Granot&Piran2012}.
In this case, the dynamics of $\Gamma(t_{\rm on})$ and the jet opening angle remain roughly unchanged after $t_{\rm jb}$.

\begin{figure*}
\centering
\includegraphics[width = 0.4\textwidth]{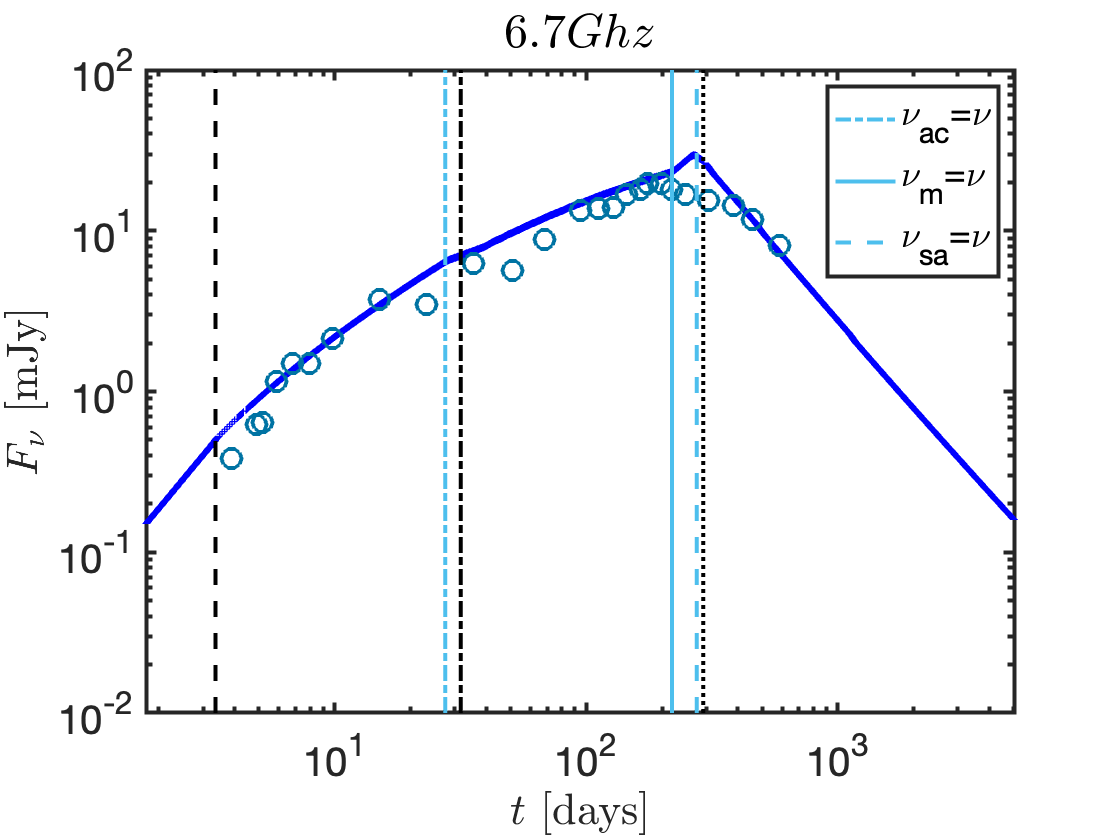}
\includegraphics[width = 0.4\textwidth]{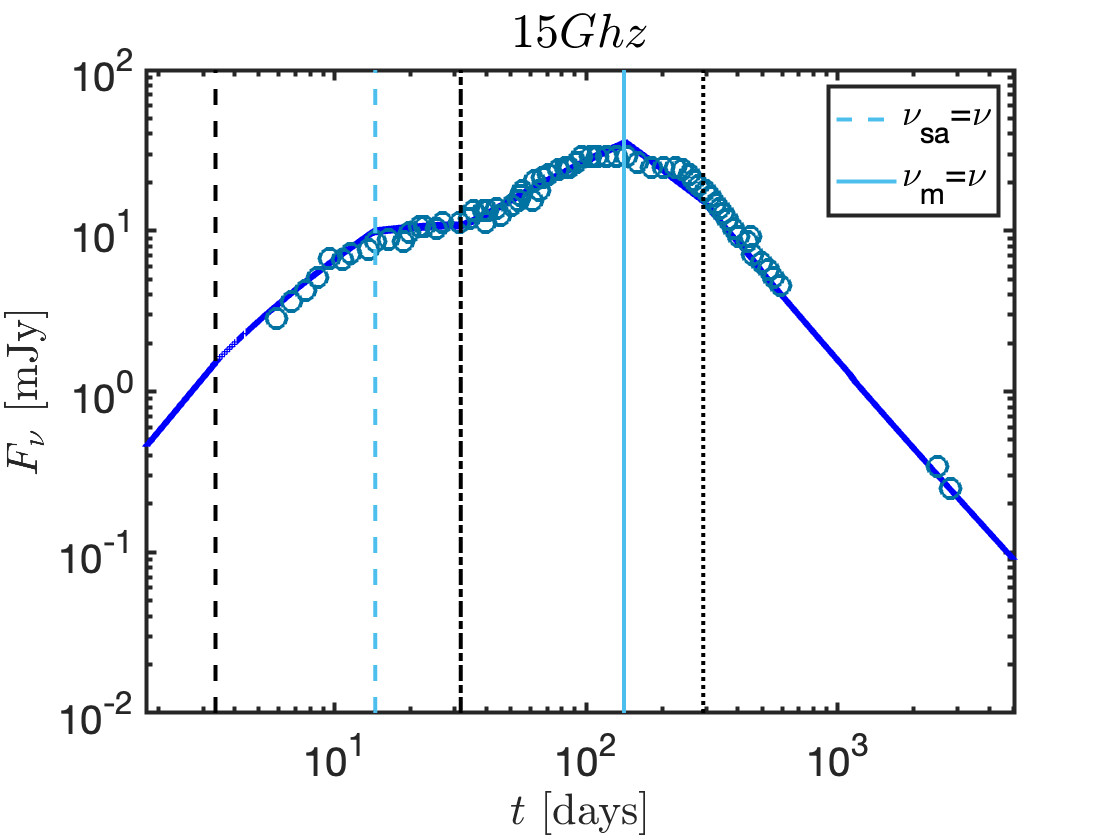}\\
\includegraphics[width = 0.4\textwidth]{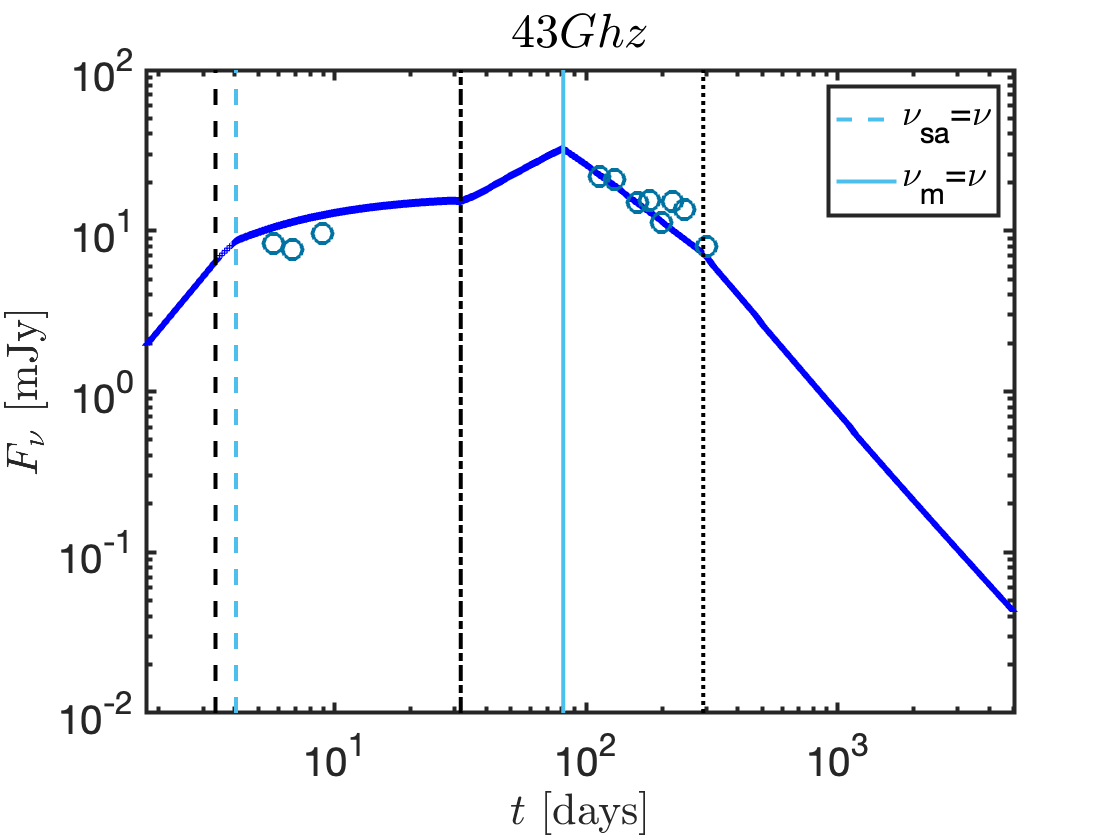}
\includegraphics[width = 0.4\textwidth]{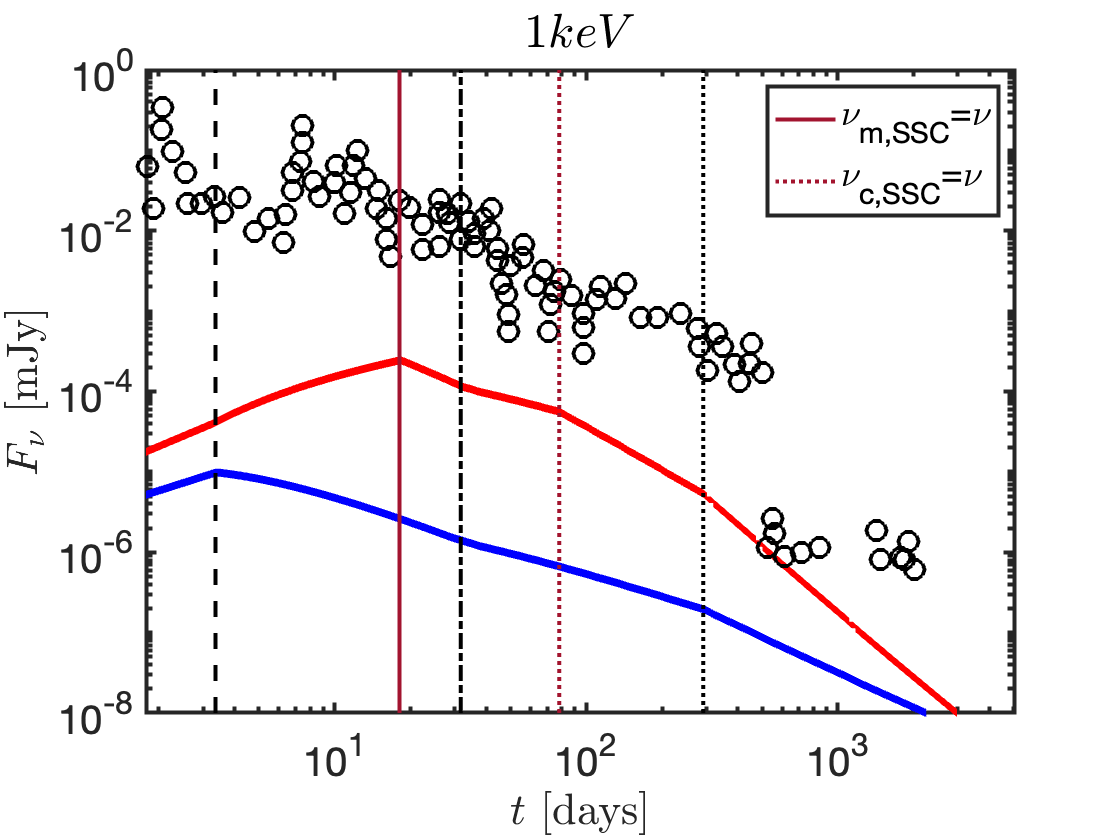}
\caption{A relativistic off-axis forward shock model for Swift J1644+57 described by $E_{\rm k,iso}=1.1\times 10^{53}\mbox{ erg},k=0.95,p=2.42,\epsilon_{\rm B}=6.3\times 10^{-3},\epsilon_{\rm e}=0.46,n(R_0)=2.9\mbox{ cm}^{-3},\Gamma_0=7,\theta_0=0.37\, {\rm rad} ,\Delta \theta=0.15\, {\rm rad}$. Different panels depict the lightcurve at different observed frequencies. Blue curves show the synchrotron contribution and the red curve show the SSC contribution. Data is shown by circles and is taken from \citealt{Zauderer2013}, \citealt{Eftekhari2018} and \citealt{Cendes2021}. Black vertical lines depict dynamical times of interest as viewed in the off-axis observer frame: $t_{\rm dec,ob}$ (dashed), $t_{\rm jb,ob}$ (dot-dashed), $t_{\rm NR,ob}$ (dotted). Other times of interest, associated with the crossing of characteristic frequencies with each other or with the observed frequency $\nu$, are shown by blue for synchrotron and red for SSC and are detailed in the legend.}
\label{fig:modelfits}
\end{figure*}

In either of the two cases above, the post-jet-break dynamics last until the outflow becomes mildly relativistic at $t_{\rm NR}\approx t_{\rm jb} \theta_0^{-2}$. At this last stage of the evolution the outflow is non-relativistic and energy conservation leads to $\beta\propto t^{\frac{k-3}{5-k}}$.
Summarizing, from the point of view of an on-axis observer, the outflow proper-velocity $u\equiv\Gamma \beta$ is estimated as
\begin{equation}
\label{eq:uevolve}
u\!\approx \!u_0 \!\left\{ \begin{array}{ll}1 & t_{\rm on}\!<\!t_{\rm dec}\ ,\\
(t_{\rm on}/t_{\rm dec})^{\frac{k-3}{8-2k}} & t_{\rm dec}\!<\!t_{\rm on}\!<\!t_{\rm jb}\ ,\\
(t_{\rm jb}/t_{\rm dec})^{\frac{k-3}{8-2k}} (t_{\rm on}/t_{\rm jb})^{-\frac{1}{2}} & t_{\rm jb}\!<\!t_{\rm on}\!<\!t_{\rm NR} ,\\
(t_{\rm jb}/t_{\rm dec})^{\frac{k-3}{8-2k}} (t_{\rm NR}/t_{\rm jb})^{-\frac{1}{2}} (t_{\rm on}/t_{\rm NR})^{\frac{k-3}{5-k}} & t_{\rm on}\!>\!t_{\rm NR}
\end{array} \right.
\end{equation}
for a jet that is maximally spreading and
\begin{equation}
\label{eq:uevolve2}
u\!\approx \!u_0 \!\left\{ \begin{array}{ll}1 & t_{\rm on}\!<\!t_{\rm dec}\ ,\\
(t_{\rm on}/t_{\rm dec})^{\frac{k-3}{8-2k}} & t_{\rm dec}\!<\!t_{\rm on}\!<\!t_{\rm NR} ,\\
(t_{\rm NR}/t_{\rm dec})^{\frac{k-3}{8-2k}}  (t_{\rm on}/t_{\rm NR})^{\frac{k-3}{5-k}} & t_{\rm on}\!>\!t_{\rm NR}
\end{array} \right.
\end{equation}
for a jet that does not start spreading while the jet is ultra-relativistic (any spreading that might occur in the non-relativistic stage, does not noticeably affect the observed lightcurve).

The number of emitting electrons at any given time/radius is:
\begin{equation}
    N_{\rm e}\approx\frac{\rho}{m_{\rm p}}\frac{\pi}{3-k}R^3 \ .
\end{equation}
We define the classical equipartition parameters $\epsilon_{\rm B}$ and $\epsilon_{\rm e}$ as the fractions of the shocked energy that are deposited in magnetic fields and relativistic electrons respectively. With these definitions, we have the magnetic field
\begin{equation}
    B\approx \left[32 \pi\epsilon_{\rm B} \rho \Gamma (\Gamma-1) \right]^{1/2}c \ , 
\end{equation}
and the minimal Lorentz factor above which electrons are accelerated to a power-law function of the form $dN/d\gamma\propto \gamma^{-p}$,
\begin{equation}
    \gamma_{\rm m}\approx \epsilon_{\rm e} \frac{p-2}{p-1}\frac{m_{\rm p}}{m_{\rm e}} (\Gamma-1) \ ,
\end{equation}
where $m_{\rm e}$ is the electron mass.
These quantities allow us to determine the synchrotron peak flux, $F_{\rm max}$ as well as the synchrotron characteristic frequencies, $\nu_{\rm m},\nu_{\rm, c},\nu_{\rm a}$ (see, e.g. \citealt{KM2008,BP2013}). We also account self-consistently  for IC losses and calculate the synchrotron self-Compton (SSC) characteristic flux and frequencies (see \citealt{PK2000,SariEsin2001}).
For the latter, we assume that for the frequencies and times of interest SSC is well described by the Thomson regime, which vastly simplifies the calculation. We verify retrospectively  that this assumption is valid for the parameters we consider.

The model described up to this point, provides us with the synchrotron and SSC fluxes at any frequency and time for an on-axis observer, $\theta_{\rm obs}\ll \theta_0$. The extension to off-axis observers can be well approximated by defining a parameter 
\begin{equation}
    a \equiv  \frac{1-\beta}{1-\beta \cos (\theta_{\rm obs}-\theta_0)} <1
\end{equation}
which is the ratio of the Doppler factors between off and on-axis observers. In terms of $a$, one finds that the flux seen by an off-axis observer, $F_{\nu}^{\rm off}(t)$, is well approximated by a re-scaling of the flux seen by an observer at $\theta_{\rm obs}=0$, $F_{\nu}^{\rm on}(t)$, using the appropriately Doppler shifted values of the time and frequency \citep{Granot2002,Kasliwal2017,Ioka2018},
\begin{equation}
\label{eq:offaxis}
    F_{\nu}^{\rm off}(t,\theta_{\rm obs})\!\approx\! F_{\nu/a}^{\rm on}(at)f_{\rm g}(\theta_{\rm obs},\theta_0)\!\left\{ \begin{array}{ll}  \!a^2 &; \theta_0\!<\!\theta_{\rm obs}\!<\!2\theta_0\ ,\\ \!(\Gamma\theta_0)^2a^3 &; \Delta \theta\!>\!\theta_0\!>\Gamma^{-1}\ ,\\ \!a^3 &; \theta_{\rm 0}\!\leq\!\Gamma^{-1}\ ,
\end{array} \right.
\end{equation}
where $\Delta \theta\equiv \theta_{\rm obs}-\theta_0$ and $f_{\rm g}$ is an order unity geometrical correction factor (see \citealt{Duque2022}).

This model enables us to  calculate self-consistently the flux at any time and frequency for an arbitrary observer as a function of 9 input physical parameters: three associated with the jet, 
$E_{\rm k,iso},\Gamma_0,\theta_0$, three with the microphysics, $p,\epsilon_{\rm B},\epsilon_{\rm e}$, two with the surrounding environment, $k,n(R_0)$ and one with the viewing angle, $\theta_{\rm obs}$.
We stress that the method of calculation presented here guarantees a self-consistent evolution of the spectral flux at different observation times. This is in contrast to equipartition analysis, in which at each observational time snapshot the characteristic flux and frequencies are fitted for independently, and as a result, their evolution as a function of time can easily be inconsistent with a fixed set of physical parameters (thus effectively increasing dramatically the number of hidden free parameters in this type of analysis) \footnote{That being said, we wish to clarify that the equipartition analysis (either the on-axis analysis or the analysis generalized to off-axis observers) is very useful in testing the viability of the underlying assumptions, and focusing our attention to any outstanding features. Indeed it was the on-axis equipartition analysis of Sw J1644 which first revealed that the available energy must be increasing with time and it is the off-axis equipartition analysis shown in \S \ref{sec:equi} that shows that an off-axis jet could in principle provide the right form of energy injection to explain the radio data.}.
Our model presented in this section can be fit to multi-wavelength, multi-epoch observations of any (initially) relativistic outflow driving a forward shock into the external medium that produces synchrotron and SSC emission. In \S \ref{sec:SwJ1644model} we apply this model to the jetted TDE, Sw J1644.

\begin{figure*}
\centering
\includegraphics[width = 1\textwidth]{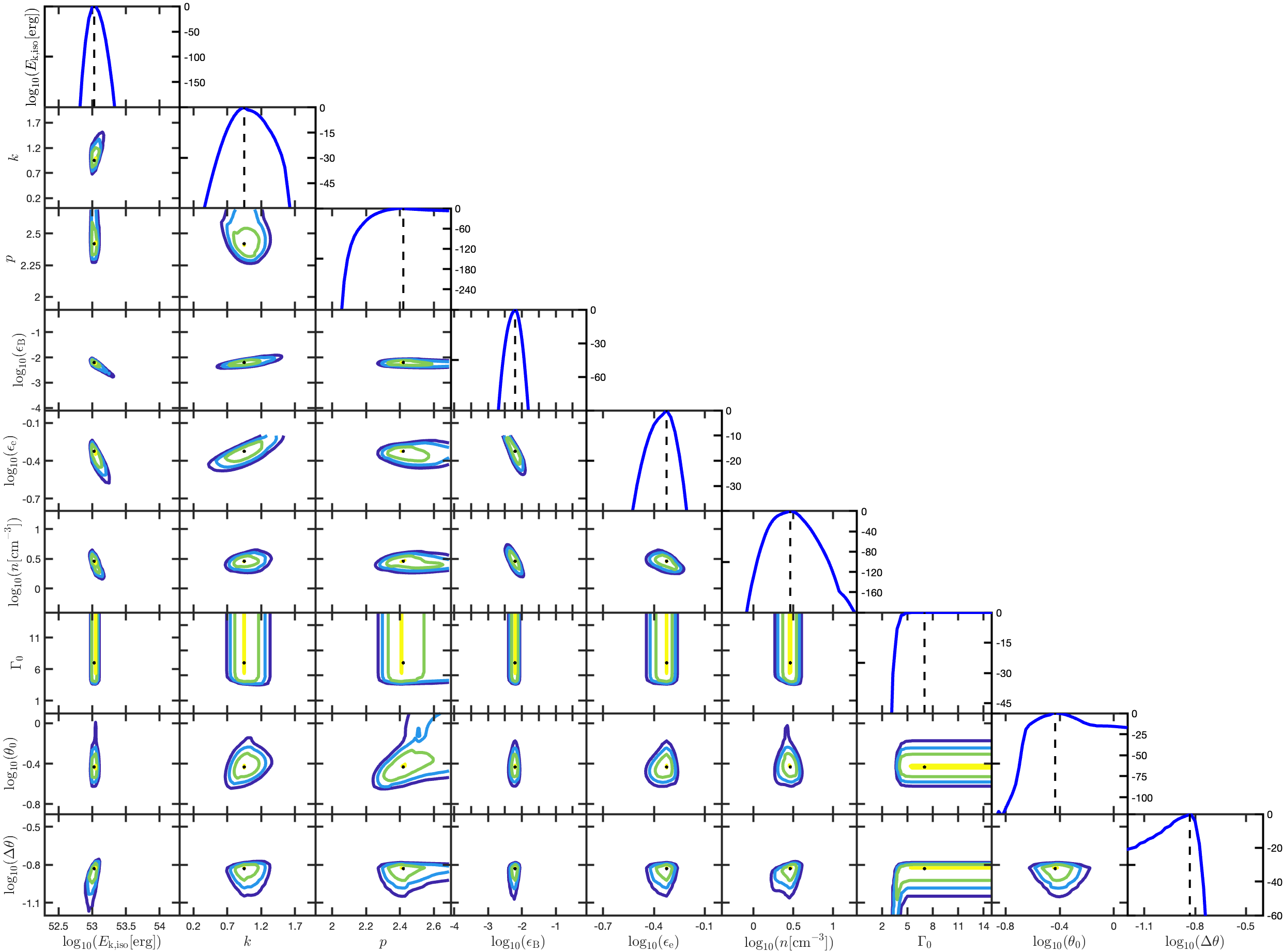}
\caption{Correlations and marginalized log-likelihood from our multi-wavelength modeling of Swift J1644+57 depicted in Fig. \ref{fig:modelfits}. Black dots and dashed lines denote our best-fit parameters. Contour lines represent marginalized likelihoods of: 0.9,1e-2,1e-4,1e-6. 
}
\label{fig:cornerplot}
\end{figure*}

\begin{figure}
\centering
\includegraphics[width = 0.4\textwidth]{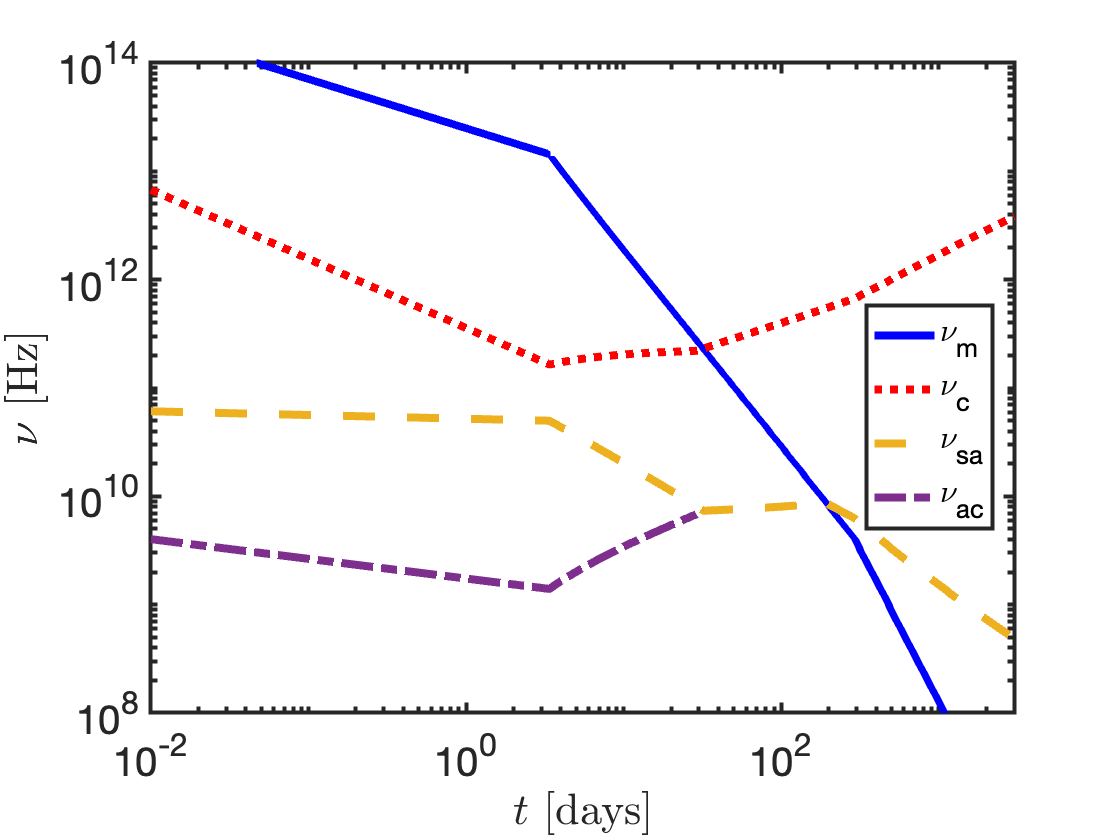}
\caption{Evolution of the synchrotron characteristic frequencies for our best-fit multi-wavelength model of Swift J1644+57 depicted in Fig. \ref{fig:modelfits} in the frame of the off-axis observer. Note that at $t\lesssim 34$\,d, the electrons are emitting synchrotron in the fast cooling regime $\nu_{\rm m}>\nu_{\rm c}$. In this regime, a new self-absorption frequency appears in the spectrum, $\nu_{\rm ac}<\nu_{\rm sa}$ since hot electrons occupy only a narrow range of radii behind the shock front (see \citealt{GPS2000}).}
\label{fig:freqevolve}
\end{figure}

\section{Modelling of TDE Sw J1644 Radio Lightcurve}
\label{sec:SwJ1644model}
We focus on the multi-band radio data of Sw J1644 reported by \cite{Zauderer2013} and \cite{Cendes2021} and perform an MCMC analysis to find the best-fit parameters for our model described in \S \ref{sec:FwdModel}. We also consider the X-ray data reported in \cite{Zauderer2013,Eftekhari2018,Cendes2021}.
The initial early X-ray lightcurve is extremely bright and highly temporally variable. At $t_{\rm j}\sim 10$\,d, it starts declining roughly as $L_{\rm X}\propto t^{-5/3}$ \citep{Burrows2011,Berger2012}. In particular, this means that most of the energy released in X-rays is emitted on a timescale of $\sim t_{\rm j}$. We, therefore, refer to this initial stage as the `prompt' X-rays, in an analogy with GRB nomenclature.
As discussed by \cite{Zauderer2013}, between $\sim 300-500$ d, the X-ray lightcurve plummets very rapidly by more than two orders of magnitude. Such a rapid decline with $\Delta t/t\ll1$ is not consistent with a forward shock and this suggests that some internal process dominates the emission at least until $t\approx 500$\,d. At later times, the X-ray lightcurve flattens. This too is not naturally explained by a forward-shock environment and indeed has been interpreted as a potential state-transition of the accretion disk \citep{Tchekhovskoy+2014}. In light of this, we opt for a conservative choice in our fitting and consider the X-ray observations only as upper-limits on the forward shock emission\footnote{We have tested fits of our off-axis forward shock model to the data including the late-time (t$\gtrsim 500$\,d) X-rays and found that these result in significantly poorer fits and are therefore disfavoured compared to the fit in which X-rays are taken as upper limits only.}. In addition, while we do not directly account for the internal process responsible for the early X-rays, we do require that the available energy towards the observer at early times should be sufficient to power the `prompt' signal.
In particular (for $\theta_{\rm obs}<2\theta_0$ as found in our best fit model described below), this leads to the constraint $E_{\rm k,iso} a(\Gamma_{10})^2\gtrsim E_{\rm X}(t_{\rm j})$, where $\Gamma_{10}$ is the Lorentz factor at an observer time of $t_{\rm j}\approx 10$\,d and  $E_{\rm X}(t_{\rm j})\approx F_{\rm X} t_{\rm j} 4\pi d_{\rm L}^2/(1+z) \approx 6.5\times 10^{52}\mbox{ erg}$ is the energy seen in the prompt X-rays \citep[we have taken the X-ray fluxes from][]{Mangano2016}.
This severely limits $\Delta \theta$.  For values of the physical parameters that are consistent with the data, this roughly corresponds to $\Gamma \Delta \theta \lesssim 2$.
An additional constraint arises from the fact that the source is unresolved by very-long-baseline interferometry \citep[VLBI, ][]{Berger2012,Yang2016}. This limits  the projected size of the source to  $R_{\rm proj}(175\mbox{ d})<0.55$\,pc, $R_{\rm proj}(3\mbox{ yr})<2$\,pc.

Applying the constraints above, our best-fit model (for the non-spreading case) is shown in Fig.~\ref{fig:modelfits} in comparison to the data. The associated 1 and 2D errors are presented in the corner plots in Fig. \ref{fig:cornerplot}.  We note that these give only a limited view of the internal correlations between model parameters, and do not preclude the existence of degeneracies between three or more of the model parameters. This is a standard limitation of corner plots when dealing with models with more than two parameters. 
The best fit parameters are: $E_{\rm k,iso}=1.1\times 10^{53}\mbox{ erg},k=0.95,p=2.42,\epsilon_{\rm B}=6.3\times 10^{-3},\epsilon_{\rm e}=0.46,n(R_0)=2.9\mbox{ cm}^{-3},\Gamma_0\geq 7,\theta_0=0.37 \,{\rm rad} ,\Delta \theta=0.15 \,{\rm rad}$. One can see that $E_{\rm k,iso},n(R_0)$ and $\Delta \theta$ are narrowly constrained. Other parameters $p$, $\theta_0$ and $\Gamma_0$ have strong lower limits. Interestingly, the obtained angles are consistent with the off-axis scenario depicted in Fig.~\ref{fig:result_j1644}, $\theta_{\rm eq}\simeq\Delta\theta+(1/\Gamma_0)\simeq0.28\,\rm rad$.
In the case of $\Gamma_0$ it is essentially unbound from above by our analysis. This is due to the fact that for our best-fit model all the observed data resides at $t\gtrsim t_{\rm dec,ob}$, at which point the value of the bulk Lorentz factor becomes independent of its initial value (see eqs. \ref{eq:uevolve}, \ref{eq:uevolve2}).
The values obtained for the equipartition parameters $\epsilon_{\rm e}$ and $\epsilon_{\rm B} $ are consistent with those found in modeling of synchrotron emission from other trans-relativistic decelerating blast waves. 
The combination of $E_{\rm k, iso}$ and $\theta_0$ enables us to estimate  the total kinetic energy of the jet $E_{\rm k} = 10^{52}$ erg. This last value is consistent with the one found from the late `on-axis' analysis of the jet (e.g. \citealt{Barniol2013A}) as at this stage the system is practically Newtonian and there are no beaming effects. 

The evolution  of the synchrotron frequencies,  $\nu_{\rm m},\nu_{\rm c},\nu_{\rm ac}$, and $\nu_{\rm sa}$ for this model are shown in Fig. \ref{fig:freqevolve}.  As shown, our model provides a good fit to the observed data.
We find that the assumption of no spreading after the jet break provides a better fit for the data.
As anticipated above, the jet is only mildly off-axis, with $\Gamma_0\geq 7$ (note that if initially $\Gamma_0\gg 7$ the jet will decelerate down to $\Gamma\approx 7$ on a time scale of a few days which is when the initial X-ray and radio observations took place) and $\Delta \theta=0.15\sim 1/\Gamma_0$. It is also noteworthy that while the synchrotron flux dominates the radio forward shock emission, the X-rays are dominated by the SSC contribution. We have verified that at the time of the X-ray observations KN corrections to the SSC flux can be reasonably ignored. 

\begin{figure}
\centering
\includegraphics[width = 0.4\textwidth]{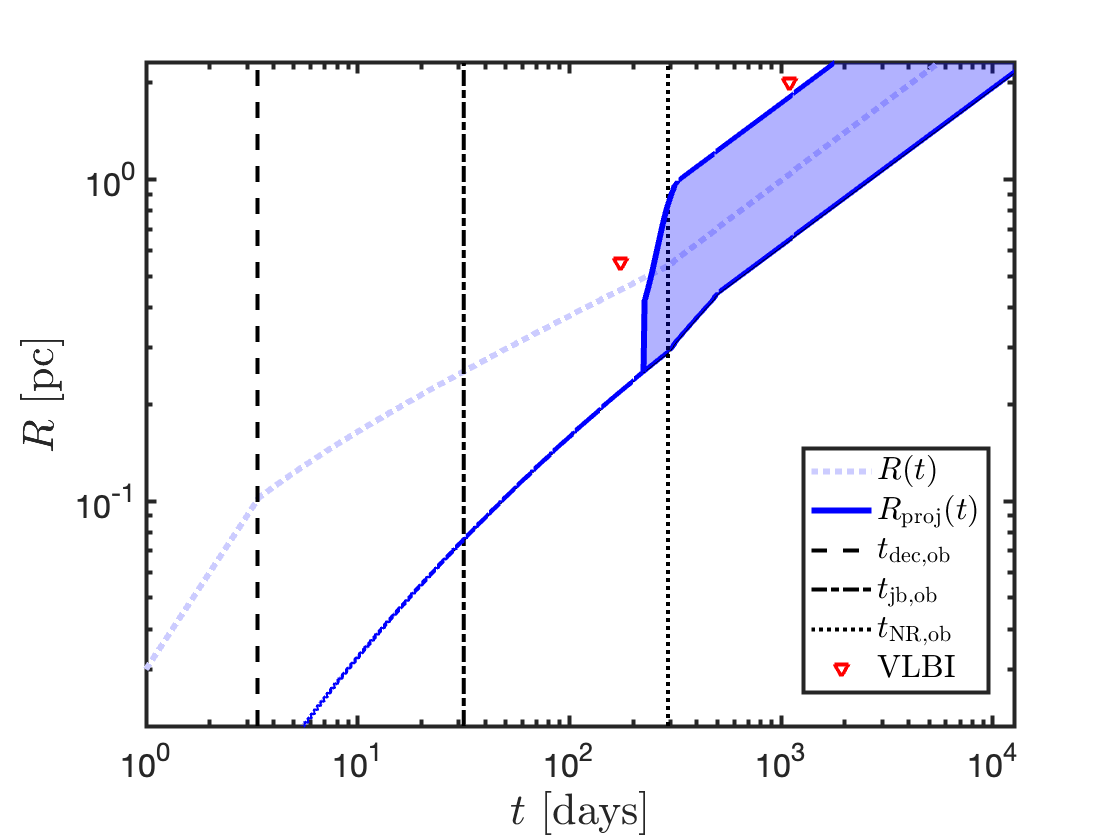}
\caption{Evolution of the emission radius and projected radius as a function of observed time for our jet model of Swift J1644+57 depicted in Fig. \ref{fig:modelfits}. Vertical lines, denote the various dynamical times described in \S \ref{sec:FwdModel}, in the frame of the off-axis observer. A shaded region depicts the range of possible values for  $R_{\rm proj}(t)$, depending on the amount of spreading that is assumed to take place once the jet becomes non-relativistic (maximal on top and none at the bottom).}
\label{fig:radius}
\end{figure}

Our model is also consistent with the VLBI limits on the projected size. For an off-axis jet, the projected size can be related to the radius of emission at the same observation time. If $2\theta_0>\max(\Delta \theta,\Gamma^{-1})$, then the whole surface of the jet, the length of which is $2R\sin \theta_0$ is effectively radiating towards the observer at the observed time. Alternatively, if $2\theta_0<\max(\Delta \theta,\Gamma^{-1})$ then the observed flux is dominated by a small portion of the jet, up to an angle of $\max(\Delta \theta,\Gamma^{-1})$ away from the line of sight, and the effective surface length becomes $2R\sin [\max(\Delta \theta,\Gamma^{-1})]$. In either case, the projection to the plane of the sky, reduces the surface length by $\cos \theta_{\rm obs}$. Putting it together, we have
\begin{equation}
\label{eq:Rproj}
    R_{\rm proj}\!=\!2 R\cos \theta_{\rm obs}    \!\left\{ \begin{array}{ll} \! \sin \theta_0 & 2\theta_0\!>\!\max(\Delta \theta,\frac{1}{\Gamma})\ ,\\
\!\sin[0.5\cdot\max(\Delta \theta,\frac{1}{\Gamma})] & \mbox{ else}
\end{array} \right.
\end{equation}
The maximal value is obtained for a spherical blast-wave, in which case the projected size is simply the diameter, $R_{\rm proj}=2R$. Note that at late times, the outflow is Newtonian or only mildly relativistic and hence whole the jet surface contributes to the emission, i.e. $2\theta_0>{\rm max}(\Delta\theta,1/\Gamma)$.
The evolution of $R(t), R_{\rm proj}(t)$ for our model are presented in figure \ref{fig:radius}. At 175 d, we find $R_{\rm proj}\approx 0.4$\,pc, and after 3 yrs, $R_{\rm proj}\approx 0.6-1.8$\,pc (the range depending on the amount of spreading that takes place once the outflow is non-relativistic), consistent with the concurrent VLBI upper limits \citep{Berger2012,Yang2016}.

\begin{figure}
\centering
\includegraphics[width = 0.4\textwidth]{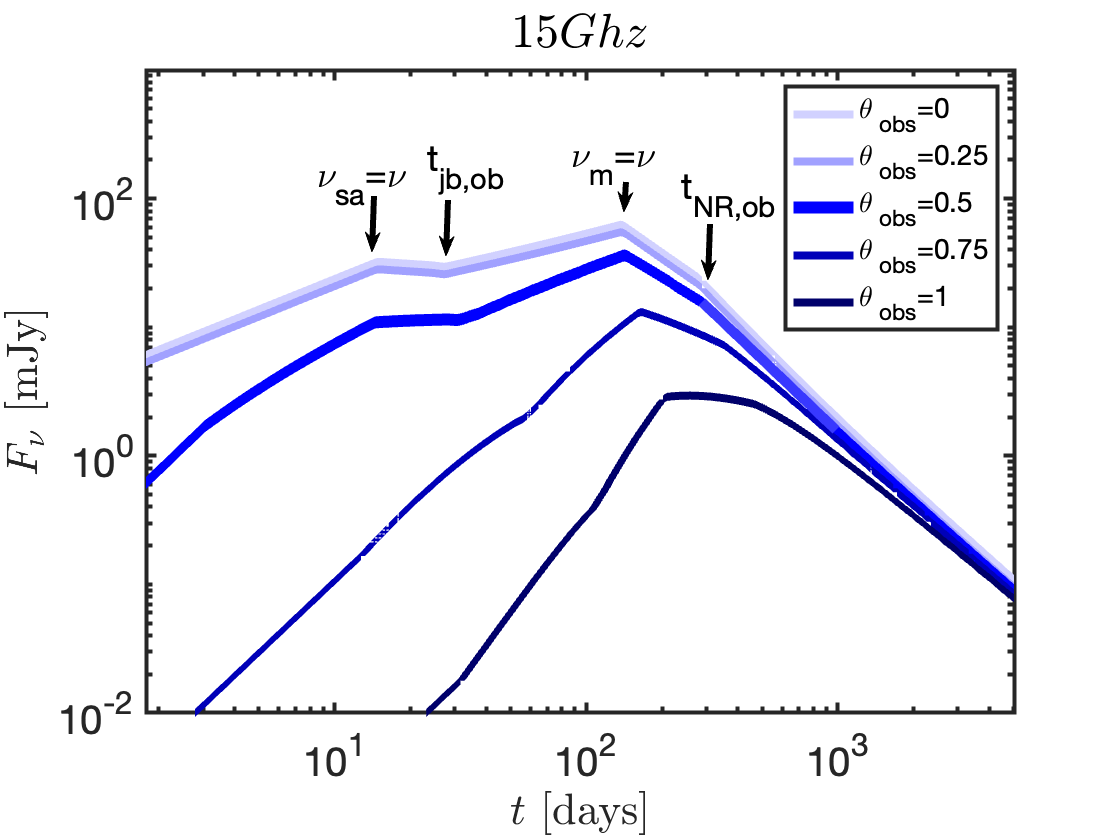}
\caption{15Ghz light-curves from a jet with the same properties as found for our best-fit model as viewed from different lines of sight towards the jet axis.
 Relevant temporal breaks are directly depicted by arrows.}
\label{fig:angles}
\end{figure}

Finally, in figure \ref{fig:angles} we show how the emission from a jet with the same properties as found in our analysis will be seen by observers looking from different viewing angles.

\section{Conclusions}
\label{sec:conc}
In this work we have constructed an analytical toy model to describe the synchrotron (and synchrotron self-Compton) emission from a relativistic off-axis jet, decelerated by an external medium with a PL density profile. We have shown here that the broad-band radio data of TDE Sw J1644, is consistent with an afterglow origin, from a relativistic jet that is directed slightly off-axis relative to our line of sight. The advantage of the approach adopted in this work is that it allows us to efficiently scan a large potential parameter space of physical conditions and involves only a minimal set of parameters that are necessary to calculate the emission self-consistently. It allows us to provide estimates of those parameters. Those could be used as a basis for more accurate, but much more resource and time consuming numerical simulations of the jet propagation and emission.

The off-axis model presented here accounts for the rising inferred energy lasting $\sim 200$\, d that is estimated from the radio flux, with no need for late energy injection by the central engine or an angular structure around the jet core (which is essential for the common on-axis model\footnote{See, however, \cite{Kumar2013}}). The off-axis interpretation adds only one parameter to the standard on-axis analysis. This should be contrasted with the on-axis energy injection interpretation, which requires at least several additional free parameters to describe increasing $E_{\rm k,iso}$ or $E_{\rm eq}$ (at the very least, the times at which energy injection begins and ends and the respective values of the energy at those times - overall four additional parameters - should be specified for that model) and further additional parameters in order to describe the change over time of the inferred slope of the power-law density profile, $k(t)$.
It is also useful to bear in mind that within the energy injection model, the time at which most of the energy is injected is $\sim 20$ times greater than the time at which the majority of the early X-ray data, commonly interpreted as internal energy dissipation in the jet, is observed (this is because after an initial phase lasting $t_{\rm j}\simeq10\,\rm days$, the X-ray luminosity begins to decline rapidly, approximately as $L_{\rm X}\propto t^{-5/3}$). 

Finally, we stress that by construction, the off-axis model presented in this work evolves in a self-consistent manner as a function of time. This is in contrast to a commonly used approach in the analysis of Sw 1644 (e.g. \citealt{Berger2012,Barniol2013A}), whereby the broadband data are fitted independently at each time snapshot, or an equipartition analysis is performed, to find the corresponding physical parameters (i.e. isotropic equivalent energy, external density, Lorentz factor, etc.), with no requirement that the temporal evolution of those parameters is self-consistent with a dynamical evolution of a jet. Moreover, the implied parameters of the synchrotron spectrum evolve inconsistently with time as compared with expectations from jet dynamics, even after accounting for multiple additional degrees of freedom that describe $E_{\rm k,iso}(t)$ and $k(t)$.  For example,  
generally $\nu_{\rm m}\propto t^{-3/2}E_{\rm k,iso}^{1/2}$. 
The on-axis modeling of \cite{Zauderer2013}, \cite{Eftekhari2018} and \cite{Cendes2021}, 
roughly results in $\nu_m\propto t^{-1}$ from 5 days and up to 2000 days after the trigger. 
As pointed out by \cite{Zauderer2013}, energy injection
can cause a flattening of $\nu_m$ (consistent with $t^{-1}$) during the first $300\mbox{ d}$, while energy is being injected.
However, once energy injection terminates at $t\sim 300$\,d the expected evolution of $\nu_{\rm m}$ should be much steeper (at least as steep as $t^{-1.5}$ and possibly steeper, if lateral spreading is important and / or once the jet becomes non-relativistic).
Note that this conclusion is independent of the radial density profile and therefore it provides a clean self-consistency test.

The best-fit parameters for the off-axis model we find here are  $E_{\rm k,iso}=1.1\times 10^{53}\mbox{ erg},k=0.95,p=2.42,\epsilon_{\rm B}=6.3\times 10^{-3},\epsilon_{\rm e}=0.46,n(R_0)=2.9\mbox{ cm}^{-3},\Gamma_0\geq 7,\theta_0=0.37 \,{\rm rad} ,\Delta \theta=0.15 \,{\rm rad}$. Such a jet has sufficient energy beamed towards the observer at early times, to explain the energy budget required to power the early X-rays. Furthermore, we have shown that the projected source size implied by this model is consistent with VLBI upper limits. The off-axis solution leads to a jet with a beaming corrected energy (assuming a double-sided jet) of $E_{\rm k,true}=7.2\times 10^{51}$\,erg, comparable to what is found in the on-axis interpretation at late-time observation, for which $E_{\rm k,true}\approx 10^{52} (\theta_0/0.1)^2$\,erg. The required energy is also consistent with the potential energy budget of a TDE.

The peak X-ray luminosity of TDE Sw J1644 is $L_{\rm X,iso}\sim 2\times 10^{48}\mbox{ erg s}^{-1}$ \citep{Bloom2011,Burrows2011,Mangano2016}. Considering the relatively large viewing angle, $\theta_{\rm obs}=0.5{\,\rm rad}$ found for the off-axis model, the true (beaming-corrected) X-ray luminosity is $\sim 2.5\times 10^{47}\mbox{erg s}^{-1}$ or $L_{\rm X,true}\sim 1200\, L_{\rm Edd}$ where $L_{\rm Edd}$ is the Eddington luminosity for $10^7\,M_{\odot}$.   Recall that the fast variability time of the prompt X-ray data \footnote{Interestingly, this fast variability suggests that the disrupted star was likely a white dwarf \citep{KrolikPiran2011}.}  $\delta t_{\rm var}\sim 100$\,s as well as the galaxy's bulge luminosity, suggest the disrupting black hole cannot be much more massive than $10^7M_{\odot}$ \citep{Bloom2011,Burrows2011}. It remains an open question how such a highly super-Eddington jet can be realized. That being said, the situation is qualitatively similar in on-axis models (unless $\theta_0\lesssim 0.03$). 

The inferred on-axis (not beaming-corrected) X-ray luminosity is relatively larger $\sim10^{49}\,\rm erg\,s^{-1}$ than other jetted TDE candidates. Nevertheless, the X-ray emitting site may not suffer from the compactness limit due to the non-detection of high energy photons $\lesssim100\,\rm keV$ \citep[e.g.,][]{Matsumoto+2019b}.

The relatively large opening angle inferred by our analysis, has implications on the intrinsic rate of TDE jets. Considering Sw J1644, if we are off-axis relative to the jet core, the isotropic equivalent energy of the jet along its core should be at least as large as the value estimated from the observed early X-ray fluence. Furthermore, the increased value of $\theta_0$ in the off-axis interpretation compared to the typical value assumed in an on-axis interpretation, suggests a larger true energy than inferred in the latter scenario. This would imply that similar TDE jets are detected more readily than previously estimated and that for a fixed detection rate we should infer a smaller intrinsic rate of such events. That being said, we stress that going beyond Sw J1644, the conclusion regarding the inferred rate of events, depends critically on the distribution of energy / opening angles between events. To simply illustrate this, we consider the case of a Euclidean geometry and in which the detectability of a jet is limited by the early X-ray fluence. Under those assumptions, jets can be detected up to a maximal radius $r_{\rm lim}\propto E_{\rm iso}^{1/2}$. The number of detected events is then approximately $N_{\rm det}\propto r_{\rm lim}^3 \theta_0^2\mathcal{R}\propto E_{\rm iso}^{3/2} \theta_0^2\mathcal{R} \propto E_{\rm true}^{3/2} \theta_0^{-1}\mathcal{R}$ (where $\mathcal{R}$ is the intrinsic rate of TDE jets) \footnote{Given the cosmological distances from which TDE jets can be detected, the detectable volume increases slower than $r_{\rm lim}^3$. This leads to deviations from the quoted power-law scaling relations, but can nonetheless be precisely accounted for analytically. We ignore this correction here for clarity.}.
For a fixed detection rate, $N_{\rm det}$, and typical collimated corrected energy between different events, we see that the inferred rate increases roughly linearly with the opening angle. Alternatively, if, for instance, the isotropic equivalent energy were to remain fixed between events, then the inferred rate would decrease with increasing opening angle, roughly as $\theta_0^{-2}$. With more off-axis events, the distributions of the energies and opening angles of TDE jets could be estimated and these ambiguities could be removed. This in turn would constrain the process of jet launching and collimation.

Considering the modest Lorentz factors inferred in TDE jets, $\Gamma\approx 3-10$, it is unavoidable that a significant sub-population of TDE jets would be detectable off-axis. Even under the most conservative assumption, which is that the jet energy effectively falls to zero beyond $\theta_0$ (a `top-hat' jet), for $\Gamma=3$ ($\Gamma=10$) and $\theta_0>0.1$ there should be up to $\sim 50$ ($\sim 7$) times more off-axis TDE jets with $E_{\rm iso}(\theta_{\rm obs})>0.01E_{\rm iso}(\theta=0)$ (such that they should remain detectable up to large distances) than on-axis ones. One can test whether the emission from a given transient is due to an off-axis jet by considering the rate at which the lightcurve rises towards the peak. 
For an off-axis jet, this temporal rise is very rapid. By measuring the spectral band in the same band (or bands) where such a rise is recorded, one can test the off-axis interpretation more quantitatively by checking the closure relation between the temporal and spectral slopes. A table with the expected values of those for all the synchrotron frequency regimes is provided in appendix \S  \ref{sec:closure}.

Going beyond a top hat jet, an angular structure extending beyond $\theta_0$ in either the kinetic energy per solid angle or the Lorentz factor could cause the fraction of off-axis to on-axis detections to increase significantly \citep{BN2019}. Such a structure could also lead to unique afterglow temporal evolutions, such as a double-peaked afterglow lightcurve (see \citealt{BGG2020}), which would be extremely unnatural to explain with an energy injection model, and which could provide conclusive evidence in favor of an off-axis interpretation. Finally, a detection of apparent superluminal motion of the flux centroid, would provide the `smoking-gun' evidence, that a given event is viewed off-axis.

\section*{Acknowledgments}
PB's research was supported by a grant (no. 2020747) from the United States-Israel Binational Science Foundation (BSF), Jerusalem, Israel. TP's research was supported by an Advanced ERC grant MultiJets. TM is supported in part by JSPS Overseas Research Fellowships.
\section*{Data Availability}
The data produced in this study will be shared on reasonable request to the authors.

\appendix
\section{Closure relations for far off-axis top hat jets}
\label{sec:closure}
\begin{figure}
\centering
\includegraphics[width = 0.4\textwidth]{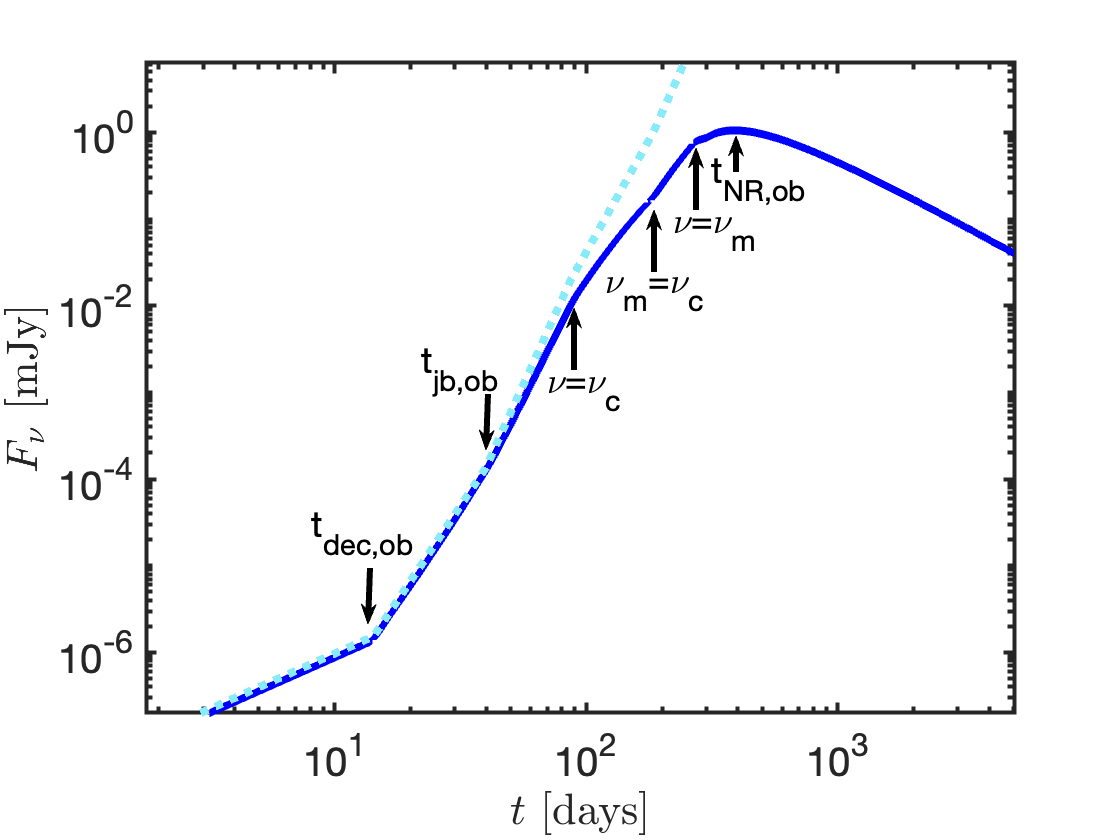}
\caption{15Ghz lightcurve of a jet with $\Gamma_0=30,\theta_0=0.1,\theta_{\rm obs}=1$ and other parameters as in our best fit for Swift J1644+57, shown in figure \ref{fig:modelfits}. A dotted line represents the asymptotic off-axis scaling expected to hold as long as $\Gamma\gg \Delta \theta^{-1}$. Indeed, the scaling matches well the numerically calculated lightcurve as it transitions from the pre-deceleration to post-deceleration and eventually post-jet break dynamical stages. The approximation breaks down at $t\approx 100$\,d, at which point $a\approx 0.25$ and the approximation that $a$ is a PL function of $\Gamma$, $a\propto \Gamma^{-2}$, breaks down.}
\label{fig:closure}
\end{figure}

Consider a `top-hat' jet that is viewed off-axis (i.e. $\Gamma\gg \Delta \theta^{-1}$). Assuming the jet dynamics are approximately unchanged after the jet break time (see \ref{sec:FwdModel}) the flux rises approximately as a power-law in time while $t<t_{\rm pk}$\footnote{In the other extreme, of maximally spreading jets, $\Gamma$ decreases roughly exponentially with radius beyond the jet break. This leads to a break-down of the power-law description of the temporal flux rise, and in turn corresponds to an even fast rise as compared with that found for a non-spreading jet.} (where $t_{\rm pk}$ is the time at which the core of the jet becomes visible to the observer, which approximately corresponds to $\Gamma(t_{\rm pk})=\Delta \theta^{-1}$ or equivalently $a\approx 1/2$). For a given power-law segment (PLS) of the synchrotron spectrum (these segments depend on the location of the observed frequency relative to the synchrotron characteristic frequencies, see \citealt{GS02}) the flux for an on-axis observer can be written as $F_{\nu}^{\rm on}\propto \nu^{-\beta_0} t^{-\alpha_0}$. 
This can be directly related to the observed off-axis flux which in this case is also characterized by power-law spectral and temporal dependencies $F_{\nu}^{\rm off}\propto \nu^{-\beta} t^{-\alpha}$ \citep{GR2010}.
Noting that for an off-axis observer, $t\propto t_{\rm on}/a$, \cite{GR2010} show that after the deceleration time $t\propto R\propto \Gamma^{-2/(3-k)}$. This leads to $a\approx (\Gamma \Delta \theta)^{-2}\propto t^{3-k}$. Using eq. \ref{eq:offaxis} with $\theta_{\rm obs}\gg \theta_0$, one finds $\alpha=\alpha_0-(3-k)(3+\beta_0-\alpha_0)$ for $\Gamma \theta_0\leq 1$ (or equivalently $t>t_{\rm jb,ob}$) and $\alpha=\alpha_0-(3-k)(2+\beta_0-\alpha_0)$ for $\Gamma \theta_0>1$ (or equivalently $t_{\rm dec,ob}<t<t_{\rm jb,ob}$).
An off-axis jet viewed from $\theta_0<\theta_{\rm obs}<2\theta_0$, evolves in the same way as a jet viewed at $\theta_{\rm obs}\gg \theta_0$ with $\Gamma\theta_0>1$.
Finally, in the case where an off-axis jet is observed at $t<t_{\rm dec,ob}$, we have $\Gamma=const$, leading to  the simple result $\alpha=\alpha_0$.
In all the cases mentioned, the spectral index remains unchanged, $\beta=\beta_0$.  
Table \ref{tab:closure} lists the corresponding values of $\alpha,\beta$ for all eight of the synchrotron PLS and figure \ref{fig:closure} shows a comparison of the analytical power-laws with a numerically calculated light-curve. For $k\approx 1, p\approx 2$, as expected for TDE jets, one finds that $3.5\lesssim |\alpha| \lesssim 7$ for $\Gamma\theta_0\leq 1$ and $1.5\lesssim |\alpha| \lesssim 5$ for $\Gamma\theta_0>1$ (or $\Delta \theta<\theta_0$), depending on the PLS. Typically, $|\alpha|$ decreases with $k$.
The closure relation between the temporal and spectral slopes can be a way to test the validity of an off-axis interpretation and potentially to also constrain $p,k$.

\begin{table}
    \centering
    \begin{tabular}{c|c|c|c|c}
    \hline
         PLS & $\beta$ & $\alpha$ ($\Gamma=const$) & $\alpha (\Gamma>\theta_0^{-1})$$^*$  & $\alpha (\Gamma\leq\theta_0^{-1})$\\
           &  & $t<t_{\rm dec,ob}$ & $t_{\rm dec,ob}<t<t_{\rm jet,ob}$  & $t>t_{\rm jet,ob}$\\
        \hline
        A & $-\frac{5}{2}$ & $-\frac{k+16}{4}$  & $k-14 \over 4$ & $5k-26\over 4$ \vspace{0.2cm}\\
        B  &$-2$ &  $-\frac{12+5k}{6}$ & $-2$ & $k-5$ \vspace{0.2cm}\\
        C  & $-\frac{11}{8}$ & $1-\frac{33k}{16}$ & -$74+3k\over 16$ & $13k-122\over 16$ \vspace{0.2cm}\\
        D  & $-\frac{1}{3}$  & $\frac{4k}{3}-3$ & $8k/3-7$ & $11k/3-10$ \vspace{0.2cm}\\
        E   & $-\frac{1}{3}$ & $2k-\frac{11}{3}$ &  $8k-17 \over 3$ & $11k-26 \over 3$ \vspace{0.2cm}\\
        F & $\frac{1}{2}$ & $\frac{3k}{4}-2$  & $9k-26\over 4$ & $13k-38\over 4$ \vspace{0.2cm}\\
        G & $\frac{p-1}{2}$ & $\frac{k(p+5)}{4}-3$  & $6(p-5)-k(p-11) \over 4$ & $6(p-7)-k(p-15) \over 4$ \vspace{0.2cm}\\
        H  & $\frac{p}{2}$ & $\frac{k(p+2)}{4}-2$ & $6p-32-k(p-10)\over 4$ & $6p-44-k(p-14)\over 4$ \vspace{0.2cm}\\
        \hline
    \end{tabular}
    \caption{Temporal and spectral slopes of the synchrotron flux (defined by $F_{\nu}^{\rm off}\propto \nu^{-\beta} t^{-\alpha}$) resulting from a non-spreading jet as viewed by a far off-axis observer. The notation of the synchrotron PLS, between A and H is taken from \citealt{GS02}. The last two columns are appropriate for jets viewed from $\theta_{\rm obs}\gg \theta_0$. For jets viewed from $\theta_0<\theta_{\rm obs}<2\theta_0$, the scaling is the same as in the second to last column, marked by an asterisk.}
    \label{tab:closure}
\end{table}

\bsp	
\label{lastpage}
\end{document}